\documentclass[aps,pre,onecolumn,longbibliography,nofootinbib,showkeys,showpacs,amssymb]{revtex4-1}

\usepackage{latexsym}
\usepackage{amsfonts}
\usepackage{color}
\usepackage{graphicx}
\usepackage{mathptmx}      
\usepackage{bm}
\usepackage{enumerate}
\usepackage{subfigure}
\usepackage{listings}
\usepackage{grffile}

\begin{document}

\title{Analysis of Wigner's Set Theoretical Proof for Bell-type inequalities\footnote{Published in J. Mod. Phys. 8, 57 (2017) + revision of Table 1}}

\author{Karl Hess}
\affiliation{Center for Advanced Study, University of Illinois, Urbana, Illinois, USA},
\author{Hans De Raedt}
\affiliation{Zernike Institute for Advanced Materials,\\
University of Groningen, Nijenborgh 4, NL-9747AG Groningen, The Netherlands}
\author{Kristel Michielsen}
\email{k.michielsen@fz-juelich.de}
\thanks{Corresponding author}
\affiliation{Institute for Advanced Simulation, J\"ulich Supercomputing Centre,\\
Forschungszentrum J\"ulich, D-52425 J\"ulich, Germany}
\affiliation{RWTH Aachen University, D-52056 Aachen, Germany}

\date{\today}

\begin{abstract}

We present a detailed analysis of the set theoretical proof of Wigner for Bell type inequalities with the following result.
Wigner introduced a crucial assumption that is not related to Einstein's local realism, but instead, without justification,
to the existence of certain joint probability measures for possible and actual measurement outcomes of Einstein-Podolsky-Rosen
(EPR) experiments. His conclusions about Einstein's local realism are, therefore, not applicable to EPR experiments and the contradiction of the experimental
outcomes to Wigner's results has no bearing on the validity of Einstein's local realism.

\end{abstract}


\maketitle

\section{Introduction}
Einstein challenged the Copenhagen interpretation of quantum mechanics by proposing with Podolsky and Rosen \cite{EPR35}
Gedanken-Experiments, briefly called EPR experiments.
These experiments were to demonstrate that quantum mechanics is
incomplete, having missed in its description of physical reality some elements of that reality.

About 30 years after the EPR paper, Bell \cite{BELL93} derived an inequality for functions of elements of physical reality,
known since as Bell's inequality, that in his opinion had to be obeyed by all of classical physics, meaning in essence by the
framework of Einstein's relativity. Wigner transformed Bell's derivation into a set theoretical approach \cite{WIGN70}.

The work of Bell and his followers had very important consequences for the views on the foundations of physics.
A majority of physicists believe that either (i) any physical
theory using counterfactually definite functions and obeying Einstein's local realism (see \ref{separation}) must obey Bell-type inequalities and/or
(ii) Wigner's derivation of Bell-type inequalities is only based on set theory (not involving counterfactual reasoning) and
Einstein's local realism. There is a considerable body of serious mathematical-physical work that has raised objections against
both (i) and (ii) (see particularly \cite{KHRE08b} and references therein as well as \cite{ROSI14} and \cite{GEUR98}). We have
shown in \cite{HESS16b} that (i) is false and it is the purpose of this paper to show that (ii) is also false.

The belief in (ii) arose mainly from the work of Wigner \cite{WIGN70} and a popularized version of Wigner's work by
d'Espagnat \cite{ESPA79}. (d'Espagnat, however, also uses counterfactual reasoning that raises additional questions
\cite{HESS16b}.) The fact that many EPR-types of experiments have been published that violate Bell's and Wigner's inequalities,
is therefore presented as proof against the validity of Einstein's local realism, because it appears inconceivable that set
theory is incorrect.

There exists, however, the following problem with this latter assessment. Set theory represents a mathematical framework based
on axioms that may be regarded as definitions. As such, set theory is not related to any measurements and experiments and,
therefore, also not to EPR experiments. It is thus necessary to find a connection between the sets of elements of physical reality
that the experiments provide and the sets of mathematical abstractions that Wigner uses in his proof. We analyze this
connection in great detail and show that Wigner's sets of mathematical abstractions contain a topological-combinatorial ordering
that is incompatible with the actual ordering of measurement-events in space and time (or space-time). We suggest that it is
this incompatibility that leads to the contradictions of Wigner's work with actual EPR measurements and not any failure of
Einstein's local realism. An analogous situation was discussed by Poincar\'e and Einstein, who resolved the experimental
contradictions to Euclidean geometry by pointing to the connection of Euclidean geometry with the physical reality, a connection
achieved by the introduction of the concept of rigid bodies \cite{EINS82}.

As we will show, there exists an analogy to the Einstein-Poincar\'e discussions, because Wigner connected his mathematical
sets to the actual measurements in a way that is inconsistent with the space-time physics of the actual EPR experiments. Key to
our reasoning is the fact that the set of data of the actual measurements is necessarily involving space- and time- (or
space-time-) coordinates to designate the correlated pairs. These sets of space and time coordinates are inconsistent with
Wigner's sets and subsets that are ordered according to location in three-dimensional space and equipment settings in that space,
but do not involve measurement times.

We show in detail that Wigner's derivation of Bell type inequalities (that restrict the possible correlations exhibited by EPR
measurements) does actually not use Einstein's local realism at all. Instead Wigner assumes in a hidden and unjustified way
certain topological combinatorial properties of his mathematical sets that are inconsistent with the macroscopic properties of
the measurement equipment in space and time as determined by clocks, protractors and meter-measures. The contradictions between
Wigner's (Bell-type) inequalities and well known experiments (see \ref{exp}) must, therefore be blamed on Wigner's
topological-combinatorial assumptions and not on Einstein's local realism.

To prove this finding, we first analyze the elements of physical reality used in EPR experiments and form the sets of physical
elements that must be dealt with in Wigner's mathematical proof (see section~\ref{separation}). We then review the essential portion of
Wigner's mathematical proof and point to his unwarranted assumptions (see section~\ref{add}). Finally, in section~\ref{sec4}
we deduce from these latter
assumptions the topological-combinatorial problems of Wigner's work and conclude that the well known line ``death by experiment"
applies to Wigner's theory not to Einstein's.
Our conclusions are summarized in section~\ref{sec5}.

\section{Einstein's realism, EPRB experiments and their resulting data} \label{separation}

We consider in the following only a variation of EPR experiments as suggested by Bohm and, therefore called EPRB experiments
\cite{ASPE82b,WEIH98}. This type of experiment involves the measurement of the two single entities of correlated pairs
of particles. Each single entity is measured at a different set of space-coordinates. These space-coordinates must, from the
viewpoint of physics, describe the entire local measurement equipment for both measurement locations. For brevity and because
important measurements have actually been performed there, we will denote the two measurement locations and corresponding space
coordinates just by ``Tenerife" and ``La Palma". The space-like separation of the measurements is a necessity to invoke
Einstein's separation principle, which is an important factor of the EPR argument and focus of the EPR discussion. Einstein's
separation principle excludes all influences between the two measurement stations that are faster than the speed of light in
vacuum.
This principle guarantees a certain independence of the two separated measurement stations and represents
the core of what is commonly called Einstein's local realism. A broader discussion and detailed definition of Einstein's local
realism has been given by Fine~\cite{Fine96}.

It is, therefore, of utmost importance for the discussions of Bell and Wigner to exclude in the actual experiments
influences from the other location that propagate with speed slower or equal to that light. How can one make sure that there exist no
such other influences and in addition that one indeed measures correlated pairs? This goal has been achieved in \cite{ASPE82b},
\cite{WEIH98} and other experiments.
These measurements relate to the Clauser-Horn-Shimony-Holt (CHSH) inequality.
We note here, however, that all of our
reasoning applies, as will become evident below, also to this type of inequality. We just discuss the Bell inequality in our
analysis, because Wigner's proof referred only to this type of inequality.

\bigskip
\subsection{Switching instruments and pairing particles}

Key to Bell's and Wigner's reasoning is the rapid switching of instruments on both Tenerife and La Palma, just before a
measurement of the spin of an incoming particle is performed. In this way one ensures that the instrument settings on the other
island cannot influence a given measurement with the speed of light or lower speeds. What Bell and Wigner wished to find out was
whether Einstein's local realism is correct or incorrect. The importance of this fast switching is emphasized over and over by
Bell \cite{BELL93} in his discussions, but no trace of it can be found in either Bell's or Wigner's formalism and algebra of
their actual proofs. The reason for this obvious lack of correspondence between theory and experiment is the assumption of Bell
and all his followers that the possible outcomes of pair-measurements are just a given for their mathematical abstractions
(functions, sets) and the fact how the pairing experimentally occurred is not included in the theory.

However, it is precisely this pairing and timing problem, where Einstein embedded the necessity of using a space-time system,
his space-time system, as the basis for his Gedanken-Experiments and the theoretical approach with Podolsky and Rosen,
\cite{EPR35}, \cite{HESS15}. There also exists no other known way to pair and correlate space like separated measurement
outcomes than by space and time measurements.%
\footnote{We do not wish to use quantum-physics arguments in our reasoning. However, it must be said that the nature of
the experiments and the possibility of quantum fluctuations excludes any other method of pairing and leaves only space and time
measurements. Suppose, for example, that we just count on each island the number of measurements and then assume that those with
equal number (deduced from the counting) will be selected as correlated pairs (post measurement, of course). Then a single
counting error (caused for example by a quantum fluctuation in the measurement equipment) will completely mess up the ordering
and will lead to pairing of the wrong entities. One must, therefore, use synchronized clocks in the space-like separated
measurement stations, as well as knowledge about the space-like separation in order to identify which detector clicks represent,
with high likelihood, measurements of a correlated pair.}

The timing of the measurements is, thus, a most important element of the physical reality, of the data, and we denote the
clock-time of the $n$th measurement in Tenerife by $t_n$ and that in La Palma by $t_n'$. Note that subsequent measurement pairs
are necessarily time-like separated because the switching of the macroscopic settings cannot be instantaneous.  To identify
actual correlated pairs, some rule needs to be employed that selects the pairs. An example of such a rule would be that the time
difference of the measurements at the two different locations (islands) must be smaller than a certain value $W$. We emphasize
that through any such rule an instantaneous logical and physical connection is established between the measurements at the two
different locations that has nothing to do with instantaneous influences, much less with information transfer. Without such a
connection, no correlations between the space-like separated measurements can be established. Only with such a connection can
intricate correlations due to physical law be revealed that now may involve also the many-body dynamics of the measurement
equipment of both locations.
It also must be emphasized that the analysis of the raw data of three different
EPRB experiments~\cite{WEIH98,AGUE09,ADEN12,VIST12}
do show a significant $W$-dependence of the amount of violations of Bell-type inequalities (actually the CHSH inequality),
for values of $W$ that are much smaller than the average time between the detection of photons~\cite{ZHAO08,RAED13a}.

From what follows in this paper, it appears imperative to perform measurements that involve dependencies on the
timing and on the instrument settings in relation to that timing.

Wigner ignored the measurement times and corresponding connections when considering possible and actual outcomes in his theory
and we will see that this negligence has serious consequences for his set theoretic calculations of correlations.

\subsection{Elements of physical reality: Notation and Notebook-entries} \label{exp}

Actual EPRB measurements (e.g. \cite{ASPE82b}, \cite{WEIH98}) monitor, in addition to the clock-times of the detector events,
two equipment orientations that determine the spin (polarization) of incoming particles. These registrations, for photons often
obtained with polarizers and avalanche photodiodes, have been denoted by a variety of symbols such as $horizontal$ and
$vertical$ etc. and Wigner uses $+$ and $-$ for any particles that he considers.
We denote the measurement result for the $n$th
measurement in Tenerife by $s_n$ and that in La Palma by $s_n'$ each being either $-$ or $+$.

As mentioned, the atoms and molecules constituting the measurement equipment have many body interactions with the incoming
particles, from both the view of Einstein's physics and also from the most modern view of quantum physics. Nevertheless, both
Bell, Wigner and all their followers describe this equipment, as the Copenhagen school of quantum mechanics does, just by a
three-dimensional unit vector of space, indicating the direction of a polarizer or of a Stern-Gerlach magnet. We denote this unit
vector for the nth experiment by ${\bf j}_n$ in Tenerife and ${\bf j}_n'$ in La Palma, respectively. Wigner assumed that this
unit vector may assume precisely the same three directions $\bf a, b, c$ in both Tenerife and La Palma and that these directions
are randomly chosen on each island in the actual experiments. Using these assumptions and notational conventions we arrive at
EPRB notebook-entry pairs of the kind:
\begin{equation} [s_n, {\bf j}_n, t_n; s_n', {\bf j}_n', t_n' ] ,
\label{25septn1}
\end{equation}
where, as mentioned, both $s_n$ and $s_n'$ may (exclusively) assume the value of either $+$ or $-$ and both ${\bf
j}_n$ and ${\bf j}_n'$ are randomly chosen from the vectors $\bf a, b, c$. {\it The parenthesis $[ \cdot]$ indicates the
measurement of a correlated pair} and the symbol ``;" denotes the separation of entries for Tenerife and La Palma. The number 9
of the time-selected pairs thus could be represented by the following actual notebook entry $[+, {\bf a}, t_9; -, {\bf b},
t_9']$. As far as EPRB experiments are concerned such entries are collections of elements of physical reality (in the sense of
Mach) and represent the only physical reality that we need to consider to discuss Wigner's proof.

As mentioned, both Bell and Wigner use Einstein locality as a tool of reasoning: For carefully designed EPRB experiments they
reason, that the measurement outcome $s_n$ does not depend on what vector ${\bf j}_n'$ is chosen in La Palma and $s_n'$ does not
depend on ${\bf j}_n$ in Tenerife.

Wigner discussed in his paper three pairs of measurements with the given settings $[{\bf a, b}], [{\bf a, c}]$ and $[{\bf b,
c}]$, exactly those that Bell had chosen for his inequality. Such a selection gives us the following six-tuple that we call
henceforth Bell's six-tuple:

\begin{equation} [s_k, {\bf a}, t_k; s_k', {\bf b}, t_k' ] \text{ }[s_l, {\bf a}, t_l; s_l', {\bf
c}, t_l' ] \text{ }[s_m, {\bf b}, t_m; s_m', {\bf c}, t_m' ] .\label{25septn2}
\end{equation}
Here $k=1,\ldots,M$, $l=M+1,\ldots,2M$, and $m=2M+1,\ldots,3M$
where $M$ is a large number determined by the total number of
measurements. The brackets $[ \cdot]$ indicates again the measurement of a correlated pair.  If we assume that these pairs are derived from
the notebook entries (\ref{25septn1}), then Einstein's local realism does not put any constraint on the possible values of the
pairs $s_k, s_k'$, $s_l, s_l'$ and $s_m, s_m'$. For given measurement times, they may assume any of the possible $2^6$
combinations of values of $+$ and $-$, because from both a physical and mathematical point of view there are at this point no restrictions to the possible outcomes except that they are two-valued.

For each pair, we are interested with Wigner in only two different possibilities when measuring on the individual particles:
either equal outcomes or unequal outcomes.
Therefore, for a string of $M$ measurements of pairs, we have $2^M$ different possible outcomes. If we have three such strings
of pairs we have $2^{3M}$ possible different correlations of the pair outcomes. Wigner reduces the number of correlations that he
considers essential significantly further, by counting only those that yield a different fraction of the final count of equal
relative to  unequal outcomes. Simple counting leads then to the conclusion that for each string of $M$ measurements of
pairs we have $M+1$ different possible outcomes that lead to a different fraction of equal and unequal outcomes
(see Table~\ref{tab1} for an illustration of the counting procedure).
Therefore we obtain for the six-tuple (2) $(M+1)^3$, possible outcomes that Wigner counts as different.%
\footnote{The mathematical proof for
this fact is easily performed by writing out the possible different outcomes in matrix-form (first column all equal, last all
different) and using the method of complete induction assuming correctness for $M$ and proving the result for $M+1$. We have
previously given a very transparent proof by using the numbers $+1$ and $-1$ for the possible equal and different pair-outcomes
respectively [8].}

\begin{table}[ht]
\caption{
Fictive example of all possible notebook entries of an experiment illustrating Wigner's method of counting.
The ``e'' (``u'') in columns $C_1$, $C_2$, and $C_3$ indicate that a pair yields outcomes that are equal (unequal)
for $M=3$ different measurements with the same setting pair and the $2^3 = 8$ possibilities given.
The last column labeled ``e/u'' gives the fraction of equal and unequal correlations $C_1$, $C_2$, and $C_3$.
The maximum number of different items in the last column is given by $M+1=4$ in this particular example.
}
\setlength{\tabcolsep}{10pt}
\begin{center}
    \begin{tabular}{ | c | c | c | c |}
    \hline
 $C_1$ & $C_2$ & $C_3$ & e/u \\
    \hline
e& e  &e  & 3/0 \\
u& u  &e  & 1/2 \\
u& u  &u  & 0/3 \\
e& u  &u  & 1/2 \\
e& e  &u  & 2/1 \\
u& e  &u  & 1/2 \\
e& u  &e  & 2/1 \\
u& e  &e  & 2/1 \\
    \hline
    \end{tabular}
\end{center}
\label{tab1}
\end{table}

As we have mentioned, Einstein's local realism has not been used up to now. We show below that it also is not used in Wigner's
further considerations that lead to his variation of Bell's inequality. Wigner's theory does thus derive, without reference to
Einstein locality, how many different correlations of equal and different outcomes may occur for the $3M$
pairs of measurements. Nevertheless, Wigner's number of possible different correlations is significantly reduced below the above
mentioned value of $(M+1)^3$ and is only proportional to $M^2$.

We ask, therefore, the question: What caused the reduction of the possible number of correlations in Wigner's proof? Our answer
is that it was not any assumption regarding Einstein's local realism. Instead, the reduction of the number of possible
correlations arises from Wigner's unwarranted mathematical assumption that is related to set theoretic probability theory.
The reduction of possible correlations is, therefore, artificial and does not apply to EPR experiments.
As mentioned, the celebrated line
``death by experiment" is correct but it applies to Wigner's theory and not, as reported in so many sensationalist articles, to Einstein's.

\section{Wigner's additional assumptions} \label{add}

Wigner did not consider any dependence of the measurement outcomes on the measurement times and just assumed that the possible
or actual outcomes $s_n$ and $s_n'$ involve automatically, so to speak per fiat, a correlated pair. The measurement locations
and times have, however, topological-combinatorial consequences for the sets of elements of physical reality that are
represented by Wigner's sets of mathematical abstractions. It is just these latter consequences that we discuss here.
\footnote{Bell's reasoning starts to deviate here from Wigner because of his introduction of counterfactually
definite functions. We have dealt with this approach in connection with a many body dynamics of particle-equipment interactions
in \cite{HESS16b}.}

Wigner defines (p. 1007, left column last paragraph of \cite{WIGN70}) ``domains of the space of the hidden variables" and claims
(in agreement with our thinking about possible outcomes outlined above) that ``we have only $2^6$ essentially different domains.
These can be characterized by symbols:
 \begin{equation} (\sigma_1, \sigma_2, \sigma_3; \tau_1, \tau_2, \tau_3)
 , \label{22octn2}
\end{equation}
all $\sigma$ and $\tau$ assuming two possible values: $+$ or $-$, and the $\sigma$ referring to the first, the
$\tau$ to the second, particle." Wigner's detailed explanations also point out that the domains characterized by
$(\sigma_1,\sigma_2, \sigma_3; \tau_1, \tau_2, \tau_3)$
correspond precisely to the respective equipment-settings
$(\bf a, b, c; a, b, c)$ which leads to the 9 possible pairings (see Eq.~(6) below)
from which the possible pairs of Bell's six-tuple (2) and possible outcomes
($s_k, s_k'$, $s_l, s_l'$ and $s_m, s_m'$ in our notation) ) can be selected.

In the next paragraph Wigner states his most crucial assumption by a definition: ``Let $(\sigma_1, \sigma_2, \sigma_3; \tau_1,
\tau_2, \tau_3)$ denote henceforth the probability that the hidden parameters assume, for the singlet state of the two spins, a
value lying in the domain which was denoted by this symbol". As innocuous as this definition looks, it contains a very big
assumption, the assumption of the existence of a joint probability for the six-tuples (\ref{22octn2}) of possible outcomes of
the measurements, while the actual measurements are only performed in pairs.

Wigner's assumption about the joint probability of the domains of variables (parameters) that determine the possible outcomes
implies the existence of consistent joint probabilities for the possible and actual outcomes of measurements such as
$(+, -, -; -, +, -)$. Note, however, that these possible outcomes are now assumed to be listed in six-tuples with the
setting-sequence of $\bf a, b, c$ on each island, in spite of the fact that actually only pairs are measured. From a set
theoretic viewpoint we must ask ourselves: Which sets is Wigner discussing? Without doubt these must be sets of measurements of
time-correlated pairs.

As an aside, Bernard d'Espagnat \cite{ESPA79} presented a counterfactual interpretation of Wigner's work by introducing
multiple hypothetical values that  every single particle ``possesses" and that are assumed to exist in addition to the possible measurement outcomes
for particle pairs. We do not make these additional unwarranted assumptions. Our reasoning below does, however, also apply to d'Espagnat's work, with the addition of our findings about
counterfactual approaches in \cite{HESS16b}.%

Wigner himself intended, as did Bell originally, to use exclusively elements of physical reality in the sense of Mach and we discuss his
work from this point of view only.

\section{Problems with Wigner's set-theoretical assumptions}\label{sec4}

As an illustration of how far reaching Wigner's assumptions are, consider Bell's six-tuple (\ref{25septn2}) which contains three
correlated pairs and add to each of the pairs an arbitrary (artificial) third notebook entry for the setting that is not already
contained in the pair-outcomes: setting $\bf c$ for the first pair, $\bf b$ for the second and $\bf a$ for the third.
As an example we obtain then for the first pair-column a triple-column:\footnote{This is precisely what d'Espagnat did
in  his variation on the theme of Wigner (see p162 of his essay in Scientific American \cite{ESPA79})}

\begin{equation} [s_k, {\bf a}, t_k; s_k', {\bf b}, t_k' ] \text{ }\rightarrow\text{ } (s_k, {\bf a}, t_k ; s_k', {\bf b}, t_k'
/ s_?,{\bf c}, t_?) .\label{24octn1}
\end{equation}
The $?$ as subscripts indicate the arbitrariness of choice of the additions.
We have separated these additions by $``/"$ instead of $``;"$, because the latter indicates the separation of the measurements
on two islands and it is not determined from which island the additions originate. The three entries are also not all correlated
in time, because $t_?$ cannot equal any of the other measurement times of the triple. Different settings must correspond to
different times in Einstein's physics. We did not use the symbol $[ \cdot ]$ on the right side of (\ref{24octn1}) but used
instead $( \cdot )$, because the pairing by measurement times cannot be guaranteed for all of them. Together with the other two
pairs of the Bell six-tuple, we obtain thus three triples each with settings $\bf a, b, c$. All the followers of Bell and
Wigner, not just d'Espagnat but also Leggett-Garg \cite{LEGG85} and others, assume explicitly or implicitly that the joint
triple probabilities exist and are common to all three triples.

However, it is a well known fact of probability theory, particularly the set theoretic probability theory of Kolmogorov, that
the existence of joint triple (and higher order) probabilities is not guaranteed \cite{KHRE08b}, \cite{HESS15}. In the above
example, each column of triples may, for example, have different joint triple probabilities and a joint triple probability that
is common to all three columns of triples may not exist. The proof of the validity of Bell type inequalities requires an
existence proof of joint triple (quadruple etc.) probabilities from the underlying physics. Wigner just assumed these
probabilities to exist and, therefore, assumed what he had to prove.

This is an enormously important point that has been overlooked also by Bell and all his followers \cite{BELL93}. If one claims
that Wigner's approach is a set theoretic approach, we need to use a set theoretic definition of probability measures. Such a
definition indeed exists and is, as mentioned, given by the probability framework of Kolmogorov \cite{FELL68}. That framework
teaches us that, for a countable number of considered possible outcomes such as $(\sigma_1, \sigma_2, \sigma_3; \tau_1, \tau_2,
\tau_3)$, one can define consistent joint probabilities if and only if the actual measurements can also be listed or taken in
form of such six-tuples so that we can establish a one to one correlation of any of Wigner's $\sigma, \tau$ with the (possible)
measurement outcomes $s_n$ and $s_n'$. Wigner has thus made a mathematical assumption of serious consequences without having any
physical or mathematical reason.

Several different cases need to be considered to present the full proof that Wigner's approach must not be applied to EPRB
experiments. We just present one example that is typical and list the following six-tuple using Wigner's subsets related to
settings $\bf a, b, c$ for both Tenerife and La Palma with a possible choice of measurement times included:

\begin{equation} (s_h, {\bf a}, t_h /
s_i, {\bf b}, t_i /s_j, {\bf c} ,t_j \text{ } ; \text{ }s_i', {\bf a}, t_i' / s_j', {\bf b}, t_j' / s_h', {\bf c}, t_h').
\label{26septn1}
\end{equation}

The symbol $``/"$ separates the three Wigner subsets on one given island. The indexes $h, i, j$,
represent natural numbers $1, 2, 3...$ that are now different from those used previously $(k, l, m)$ in the six-tuple
(\ref{25septn2}). The reason for using a different notation is that six-tuples (\ref{25septn2}) and (\ref{26septn1}) cannot be
transformed into each other without violating the space and time correlations of the actual pairing arising from the original
measurements. The particular chosen pairing of six-tuple (\ref{26septn1}) is determined by the measurement times and thus is
given by $[{\bf a; c}]$, $[{\bf b; a}]$ and $[{\bf c; b}]$, while the pairing of six-tuple (\ref{25septn2}) is $[{\bf a; b}]$,
$[{\bf a; c}]$ and $[{\bf b; c}]$ and the pairing by measurement times must be different whenever the topology (location and setting) of the
measurements of the pairs is different.

Wigner never included considerations of both measurement times and topology and used six-tuple (\ref{26septn1}) to
obtain the possible outcomes for all the following 9 possible pairings:
\begin{equation} ({\bf a; a}), ({\bf a; b}), [{\bf a;
c}], [{\bf b; a}], ({\bf b; b}), ({\bf b; c}), ({\bf c; a}), [{\bf c; b}], ({\bf c; c}). \label{22octn1}
\end{equation}
We use again $[ \cdot ]$ only for the guaranteed correlated pairs (through measurement times).
For all other pairs we use $( \cdot )$.
For the pairings of Bell's six-tuple (\ref{25septn2}) and using (\ref{22octn2}) Wigner thus obtains:

\begin{equation} (\sigma_1,
{\bf a}; \tau_2, {\bf b}) \text{ }[\sigma_1, {\bf a}; \tau_3, {\bf c} ] \text{ }(\sigma_2, {\bf b}; \tau_3, {\bf
c}) .\label{22octn3}
\end{equation}

The $\sigma$ and $\tau$ denote here a certain given outcome of either$+$ or $-$. For
example, we may have $\sigma_1 = +$ , $\tau_3 = -$, $\tau_2 = +$ and $\sigma_2 = -$. This innocently looking fact represents a
big restriction for the possible outcomes. The two pairs with settings $(\bf a; b)$ and $[{\bf a; c}]$ must now have the same
outcome $\sigma_1=+$ in Tenerife and the same is true for the two pairs $[{\bf a; c}]$ and $({\bf b; c})$ which now must have
identical outcome $\tau_3= -$ in La Palma in spite of the fact that they must, in principle, correspond to two different pairs
of actual measurement.

We consider now the possible correlations for the outcomes of the $M$ six-tuples of Bell (corresponding to $3M$ measurements of pairs) that lead to different ratios of equal and different outcomes of the pairs.
The number of these correlations is reduced from the original $(M+1)^3$ to only $2(M+1)^2$ different possibilities of correlations as can be seen from six-tuple (\ref{22octn3}). If we also
require that $\tau_2 = \sigma_2$ as Wigner did, because he argued that equal settings need to have equal outcomes, then we
obtain only $(M+1)^2$ different possibilities of correlating equal $++,--$ and different $+-,-+$ outcomes for the pairs in
Bell's six-tuples (\ref{25septn2}). This latter reduction is identical to the reduction obtained by Bell in his inequality.

As seen from the measurement times, only 3 of the pairings of (\ref{22octn1}) do correspond to originally correlated pairs.
Note that $(\bf a; b)$ and $[\bf b; a]$ cannot be treated on the same footing, although they have only interchanged settings. It
is also very important to note that only one pair of Wigner's (\ref{22octn3}) is actually correlated. Each of the equal settings
chosen on the same island must appear with different measurement time in Bell's six-tuple (\ref{25septn2}). In contrast, any
setting that appears twice on the same island must be associated with the same measurement and, therefore, be related to the
same measurement time because of Wigner's procedure involving six-tuple (\ref{26septn1}) . {\it However, it is physically
impossible to actually measure two different pairs which exhibit the same setting and the same measurement time at one given
location}.

The ordering of the mathematical abstractions that form the subsets that Wigner chose are simply incompatible with
the ordering of the original measurements in space and time. The measurement time is the only guarantee for the correct pairing
of the original measurements. These facts demonstrate that Wigner oversimplified complex topological-combinatorial factors that
are vital for the outcomes of his considerations. Another important fact, for the incorrectness of Wigner's pairings that arise
from his 9 possibilities (\ref{22octn1}), is that the pairs with equal settings are not derived from originally correlated pairs
and therefore may randomly assume all of the value-pairs $++$, $--$, $+-$ and $-+$. They contribute to the equal outcome
$++, --$ count, that is so important for Wigner's reasoning, only half of what Wigner actually assumed.

Thus, Wigner reduced the number of possible correlations of EPRB measurements by assuming without justification the
existence of certain joint probabilities. He did not realize this fact and was, therefore, faced with the problem to explain
the contradictions with actual experiments if they occurred
(and as indeed were found~\cite{ASPE82b,WEIH98,AGUE09,ADEN12,VIST12}).
Because Wigner was convinced that he used only set theory and Einstein's local realism, he mentioned that a violation of Einstein's
separation principle would lead to ``$4^9$ domains .....of the nine measurements" corresponding to Eq.~( \ref{22octn1}). He thus
introduced violations of Einstein's realism to explain possible future disagreements of his theory with experiments and
measurements, so to speak as a deus ex machina, that would resolve contradictions; not as essential part of his proof.

We note in passing that all of our (and Wigner's) reasoning that is related to the triple of settings $\bf a, b, c$ and Bell's
inequality can be repeated with the same findings for quadruples $\bf a, b, c, d$ and the CHSH inequality.

\section{Conclusion}\label{sec5}

These facts show that Wigner's procedure to derive Bell's inequality is set-theoretically neither general nor sound.
Wigner's work, taken in conjunction with experiments such as presented in \cite{ASPE82b,WEIH98,AGUE09,ADEN12,VIST12} does not prove
violations of Einstein's realism. It proves only that Wigner's assumptions about the existence of certain joint probabilities
are incorrect.  As discussed above, the essential part of Wigner's proof does not use Einstein's local realism at all.
Thus, we believe to have shown beyond any reasonable doubt that Wigner derived his
reduction of possible correlations and his Bell type inequality from set theoretically unjustified assumptions about the
existence of joint probabilities.

\section*{Aknowledgement}
The authors thank Michael Revzen for bringing to our attention that the caption of Table I was lacking clarity.

\end{document}